\documentclass{PoS}

\usepackage[utf8]{inputenc}
\usepackage{graphicx}
\usepackage{subfigure}
\usepackage{amsmath,amssymb}%,amsthm,amsopn,mathrsfs}

\title{Screening in two-dimensional gauge theories}
\ShortTitle{Screening in two-dimensional gauge theories}

\author{Piotr Korcyl,$^{ab}$ \speaker{Mateusz Koren}$^{a}$\\
\llap{$^a$}M. Smoluchowski Institute of Physics, Jagiellonian University,\\
Reymonta 4, 30-059 Cracow, Poland\\
\llap{$^b$}NIC, DESY Zeuthen, Platanenallee 6, D-15738 Zeuthen, Germany\\

E-mail: \email{piotr.korcyl@desy.de}, \email{mateusz.koren@uj.edu.pl}}

\abstract{We analyze the problem of screening in 1+1 dimensional gauge theories. Using QED$_2$ as a warm-up for the
non-abelian models we show the mechanism of the string breaking, in particular the vanishing overlap of the Wilson loops
to the broken-string ground state that has been conjectured in higher-dimensional analyses. We attempt to extend our
analysis to non-integer charges in the quenched and unquenched cases, in pursuit of the numerical check of a renowned
result for the string tension between arbitrarily-charged fermions in the massive Schwinger model.}

\FullConference{The 30 International Symposium on Lattice Field Theory\\
Cairns, Australia\\
June 24-29, 2012}

\begin{document}

\section{Introduction}

Two-dimensional $U(1)$ gauge theory with fermions (QED$_2$) has long been a test-bed for concepts relating to
four-dimensional gauge theories. Being extremely simple compared to QCD it captures some of its crucial non-perturbative
features such as confinement and string breaking. Although over the years there has been a plethora of numerical studies
of QED$_2$ using various methods such as discrete light cone quantization, hamiltonian lattice field theory, euclidean
LFT (see Refs.~\cite{DLCQ,CKS12,GHL99} and references therein), this system still attracts a lot of attention, see e.g.
Refs.~\cite{CKS12,NAR12}.

As far as the pure gauge theory (Quantum Maxwell Dynamics, QMD$_2$) is concerned, when formulated in $\mathbb{R}^2$ it
turns out to be trivial and results in a linear confining potential for the probe charges \cite{ROT05}. On a torus the
situation is altered -- due to the fact that one cannot gauge away the fields along closed contours there exists a
single space-independent quantum degree of freedom and the spectrum of the theory yields the Manton's model
of QMD$_2$ on a circle \cite{MAN85} in the continuum limit \cite{KKW12}.

The massless case for $N_f=1$ was solved by Schwinger \cite{SCH62} using bosonisation trick, that however is not easily
generalized to the massive or multiflavour case. The solution shows that all real values of external charge
$Q_\text{ext}$ are screened by the vacuum polarization\footnote{In this letter we always write $Q_\text{ext}$ as a
dimensionless quantity, i.e. the multiplicity of the fundamental electric charge.}. Subsequently, a perturbative
addition of a small fermion mass was introduced by Coleman et al. in Ref.~\cite{CJS75} -- it was shown that with
$m\neq0$ only the integer probe charges are screened by fermion-antifermion pairs, in a mechanism reminiscent of the
string breaking in QCD. The non-integer charges are expected to have a non-vanishing string tension in accordance with
\begin{equation}
\sigma \sim m \left(1-\cos(2\pi Q_\text{ext})\right).
\label{eq:cos}
\end{equation}

Non-abelian two-dimensional models, which are the ultimate aim of our investigation \cite{KK11} are even more
interesting. There are theoretical predictions for the spectrum of QCD$_2$ with fundamental fermions \cite{QCD2} and the
large-$N$ limit of the theory was analytically solved by 't Hooft \cite{THO74}. The practicality of this solution is
however limited by the fact that the fundamental matter is quenched in the large-$N$ limit, thus even more interesting
are the large-$N$ limits of theories with two-index representation matter where the fermion dynamics plays an important
role.

For example, the adjoint fermions in 1+1 dimensions were analyzed both theoretically, giving encouraging results
\cite{KUT94}, and numerically by DLCQ techniques \cite{BHA93}. However, no lattice analysis has been performed until
recently.

In this letter we report the results of our study of QED$_2$, treated as a warm-up for the non-abelian models. In
Section \ref{sec:str} we study the string breaking in QED$_2$ using Wilson loops and in Section \ref{sec:frac} we sum up
our attempts to define operators carrying fractional charges and to calculate the string tension with them, which would
lead to a lattice verification of Eq.~(\ref{eq:cos}).

\section{String breaking in QED$_2$}
\label{sec:str}

We analyze lattice QED$_2$ with $N_f=1$ (the massive Schwinger model) by means of euclidean Monte Carlo simulations. We
use Wilson fermions and generate the configurations using the Rational Hybrid Monte Carlo algorithm. As a test of the
possibilities of the simulation we analyze the string breaking using Wilson loops.

It is well known that QED$_2$ resembles QCD$_4$ in this respect. It exhibits a confining linear potential in the pure
gauge case (as does pure gauge lattice QCD$_4$ \cite{CRE83}) and charge screening at large distances, interpreted as
string breaking by quark-antiquark pair \cite{CJS75,DRU98}. The string breaking is very
hard to observe in the lattice QCD using Wilson loops alone -- a poor overlap of the Wilson loop to the broken-string
ground state was postulated and a larger set of operators had to be used to deliver a firm observation \cite{BAL05}.

\begin{figure}[tbp!]
\subfigure[Extracted energies $E_i$]{\label{fig:wilson1}\includegraphics[width=7.25cm]{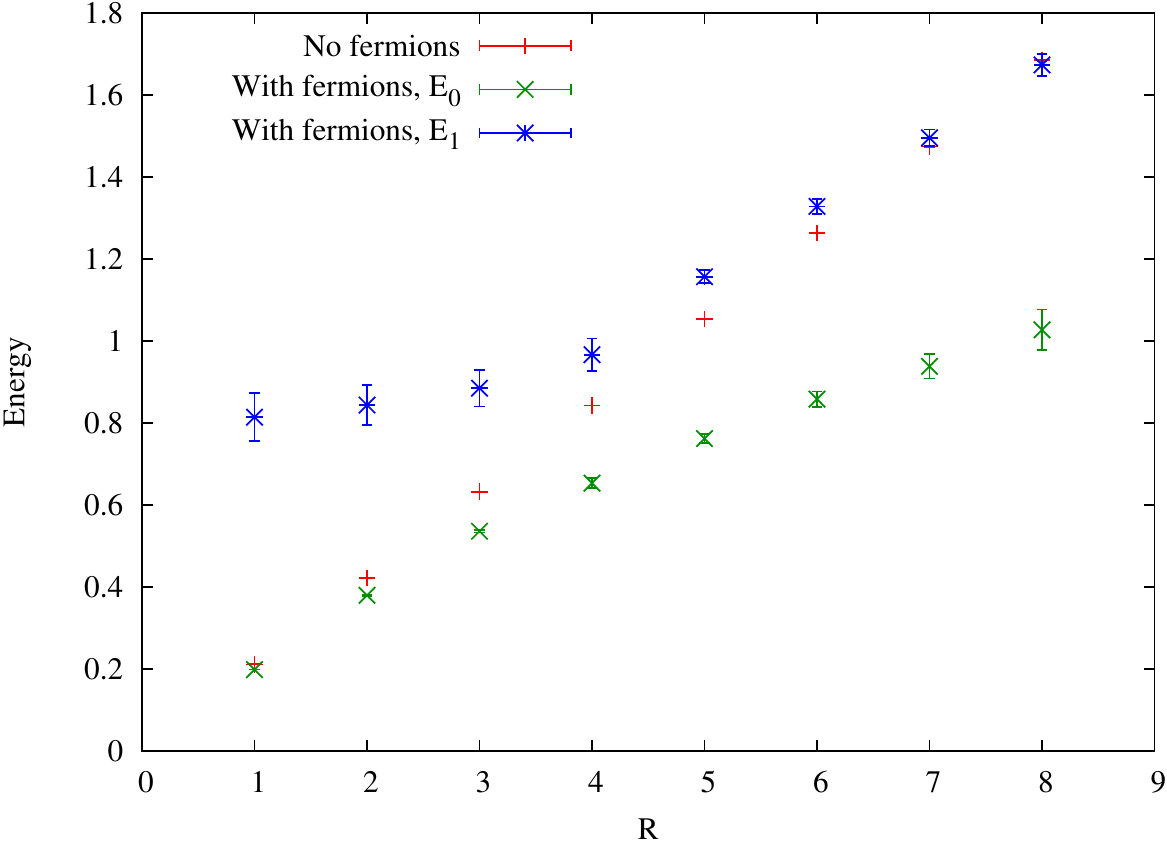}}
\hspace{0.25cm}
\subfigure[Extracted overlaps $C_i$]{\label{fig:wilson2} \includegraphics[width=7.25cm]{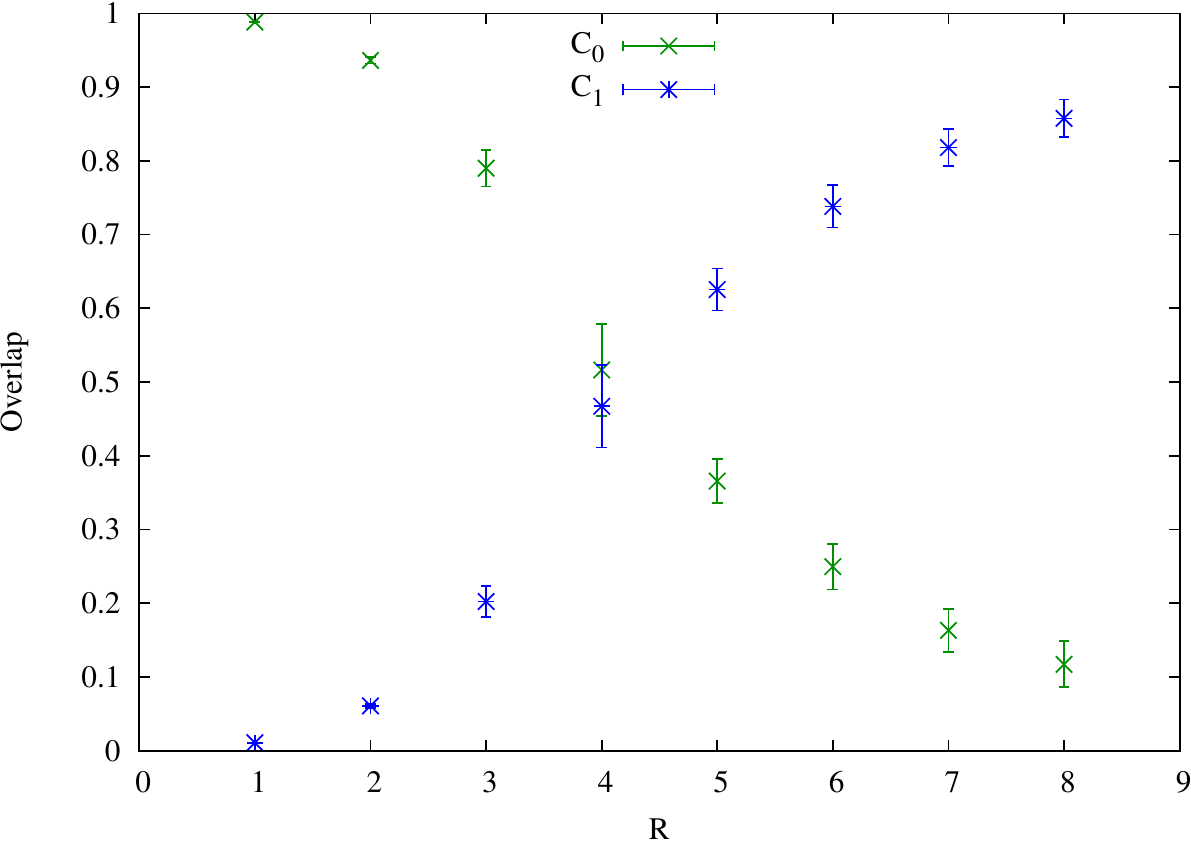}}
\caption{Energies and overlaps to the Wilson loop of the ground state and first excited state, $V=24\times24$,
$g^{-2}=1.5$, $\kappa=0.245$. The red points of in Fig.~(a), inserted for comparison, are the numerical calculation of
the (analytically known) pure gauge result.}
\label{fig:wilson}
\end{figure}

As QED$_2$ is computationally much less demanding than QCD$_4$ we are able to use a different approach. By using very
high statistics (tens of millions configurations) we are able to perform two-exponent fits to the unsmeared Wilson loop
data
\begin{equation}
W(R,T) \cong C_0e^{-E_0(R)T}+C_1e^{-E_1(R)T},
\end{equation}
extracting the ground state and the first excited state from the Wilson loop together with the corresponding overlaps.
Sample results are presented in Fig.~\ref{fig:wilson}. It is natural to interpret Fig.~\ref{fig:wilson1} as a confined
state and a broken-string state, where the latter becomes the ground state around $R\simeq4-5$, with quantum repulsion
(i.e. mixing) of states observed. Perhaps the most instructive result of this exercise it Fig.~\ref{fig:wilson2} where
one can see how the Wilson loop clearly prefers the confined state, explaining the difficulty of observing the string
breaking using solely this operator.

\section{Non-integer charges}
\label{sec:frac}

In this section we analyze the possibility of introducing operators carrying arbitrary real charge $Q_\text{ext}$ to
verify the equation (\ref{eq:cos}) (from now on, to be concise, we refer to $Q_\text{ext}$ simply as $Q$). This is a
relatively unexplored topic on the lattice. One notable exception is Ref.~\cite{HAM82} presenting a hamiltonian lattice
analysis where real charges were introduced by means of constant electric field. In this work we have approached a
different method. We define the ``charged Wilson loop'' over a contour $\Gamma$ as
\begin{equation}
W_Q(\Gamma) = \prod_{j\in\Gamma} \left(U_j\right)^Q.
\label{eq:fact}
\end{equation}

Note that one can formulate an alternative definition as
\begin{equation}
W^\text{(alt)}_Q(\Gamma) = \Big(\prod_{j\in\Gamma}U_j\Big)^Q.
\label{eq:nfact}
\end{equation}
While the two coincide for integer charges, they can give vastly different results for the non-integer case when the
complex logarithms used to define the non-integer powers fall on different branches. The numerical results presented in
this work are obtained using the first definition (\ref{eq:fact}). We will briefly discuss the different results given
by the second method later in this section.

To find the string tension dependence on $Q$ in the massive Schwinger model we have performed similar analysis as for
the ``ordinary'' Wilson loops in the previous section. For every real $Q$ we found that up to additional perimeter
terms, that do not influence the string tension, the Wilson loops with charge $Q$ behave in a very similar manner to the
ones with the integer charge closest to $Q$ -- thus for every charge analyzed the data is consistent with $\sigma_Q=0$.

To understand this result we went to an even simpler theory i.e. the pure gauge model (QMD$_2$). There we know the exact
results for the string tension of integer charges both on infinite \cite{ROT05} and finite lattices \cite{KKW12}. In the
continuum limit one might also expect the well-known continuum result $\sigma_Q\sim Q^2/2$ for all real charges.

\begin{figure}[tbp!]
\subfigure[String tension as a function of $Q$]{\label{fig:nonint1} \includegraphics[width=7.25cm]{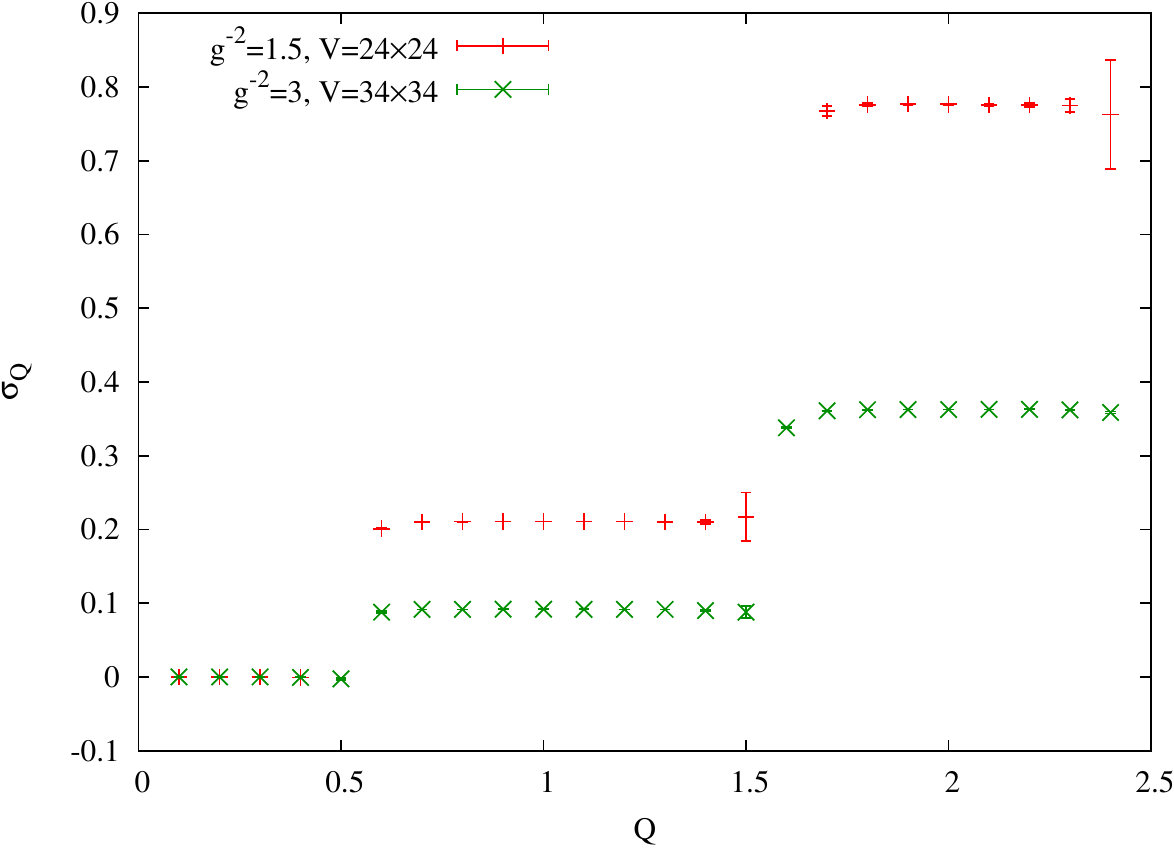}}
\hspace{0.25cm}
\subfigure[Ansatz vs. data (not a fit)]{\label{fig:nonint2} \includegraphics[width=7.25cm]{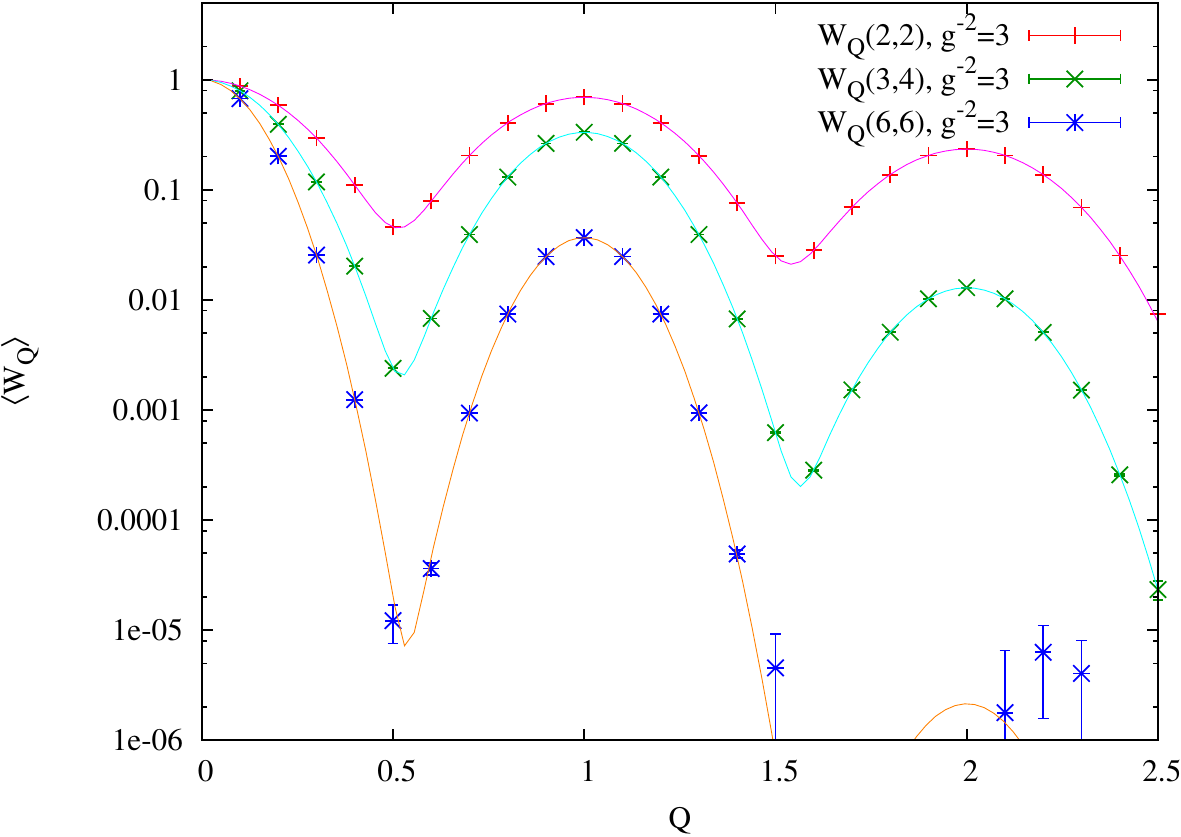}}
\caption{Charged Wilson loops in QMD$_2$.}
\label{fig:nonint}
\end{figure}

One might thus expect that $\sigma_Q$ for non-integer charges smoothly interpolate between the analytically known
results for integer values. However, the values of the string tension obtained from the single-exponent fits (see
Fig.~\ref{fig:nonint1}) show that the string tension is projected to the nearest integer-charged value\footnote{This
result shows no sign of volume dependence, up to the largest analysed lattice of volume $256\times256$.}. A similar
behaviour is observed in the pure gauge case of Manton's model of continuum QED$_2$ on a spatial circle \cite{MAN85}. In
this model one can set the gauge so that $A_x(x,t)$ is independent of $x$ and that $A_x\in[0,1)$ with both ends
identified -- the periodicity of the physical space results in the periodicity of the field space. One then obtains a
quantum-mechanical system with a hamiltonian
\begin{equation}
H=\frac{\pi}{e^2}\dot{A}_x^2
\end{equation}
and the Hilbert space consisting of wave functions satisfying $\psi(A_x=0) = \psi(A_x=1)$. Only integer-charged states
satisfy periodicity. A state created by an operator with arbitrary charge $Q$ is a result of projection onto the
physical Hilbert space. To see this let us define spatial Polyakov loop in this model as $P(A_x,\tau) = e^{2\pi i
A_x(\tau)}$. The ``charged'' Polyakov loop is then simply:
\begin{equation}
P_Q(A_x,\tau) = e^{2\pi i Q A_x(\tau)}.
\end{equation}
The correlation function of the Polyakov loops is then a sum over integer-charged states with proper overlaps:
\begin{equation}
\langle P^\dagger_Q(A_x,\tau)P_Q(A_x,0)\rangle = \sum_{n=0}^\infty e^{-n^2e^2\pi\tau} \left(\frac{\sin(\pi
  (Q-n))}{\pi(Q-n)}\right)^{2}.
\end{equation}
Driven by this analogy, we formulated an ansatz that very well describes the data (see Fig.~\ref{fig:nonint2}):
\begin{equation}
W_Q(R,T) = \sum_{n=0}^{\infty}
  \Big(\frac{I_n(\beta)}{I_0(\beta)}\Big)^{RT} \Big(\frac{\sin(\pi (Q-n))}{\pi(Q-n)}\Big)^{2R+2T}.
\end{equation}
The ansatz assumes that the finite size effects for the integer charges are negligible. In fact, using the results of
Ref.~\cite{KKW12} one can include them, leading to an exact result \cite{KKW13}.

The results obtained using the second definition, Eq.~(\ref{eq:nfact}) lead to similar conclusions with a similarly
successful ansatz of the form
\begin{equation}
W^\text{(alt)}_Q(R,T) = \sum_{n=0}^{\infty}
  \Big(\frac{I_n(\beta)}{I_0(\beta)}\Big)^{RT} \Big(\frac{\sin(\pi (Q-n))}{\pi(Q-n)}\Big),
\end{equation}
the only difference being the exponent of the sine part\footnote{Also calculations using Polyakov loops defined with
both methods lead to the same conclusions and differ from the Wilson loops only by the exponents over the sine parts.}.

The periodicity of the configuration space in the compact formulation implies that the operator with non-integer charge,
which does not satisfy periodicity
\begin{equation}
e^{iQ(A+2\pi)}\neq e^{i Q A},
\end{equation}
creates a state which is a combination of states belonging to the Hilbert space, i.e. the integer-charged states. We
conjecture that the same phenomenon occurs for dynamical fermions and is the reason of the incompatibility of our string
tension calculations and Eq.~(\ref{eq:cos}). When only the integer part of the probe charge has a physical meaning then
the effects of fractional charges are projected out, which explains the vanishing of the string tension for all real
$Q$.

\section{Conclusions}

In this work we have studied the phenomena of confinement and screening in two-dimensional lattice gauge theories. Using
high precision data from Wilson loops we investigated the string breaking in QED$_2$ and found a rapidly decreasing
overlap of Wilson loops on the broken-string state, which accounts for the difficulties in observing the string breaking
in lattice QCD simulations which use solely this operator.

Then we proposed two generalizations of Wilson loops for arbitrary real charges $Q$. We found that for both of them the
string tension vanishes for any $Q$, which would be in disagreement with the prediction of Coleman et al.~\cite{CJS75}.
Based on the experience from the pure gauge theory, we suggested an explanation by pointing out that in the compact
formulation the effects of fractional charges are projected out of the Hilbert space and only the integer charges have a
physical meaning. The value for the string tension between integer charges predicted by Eq.~(\ref{eq:cos}) is indeed
zero for all values of the fermion mass.

\section*{Acknowledgments}

This work was partially supported through NCN grant nr 2011/03/D/ST2/01932, and by Foundation for Polish Science MPD
Programme co-financed by the European Regional Development Fund, agreement no. MPD/2009/6. Numerical simulations were
performed on \verb#deszno# supercomputer at the Faculty of Physics, Astronomy and Applied Computer Science, Jagiellonian
University, Cracow.

\end{document}